\documentclass{optica-article}

\journal{opticajournal} 

\articletype{Research Article}

\usepackage{lineno}

\newcommand{\dd}[1]{\frac{d#1}{(2\pi)}}
\newcommand{\ddb}[1]{\frac{d^2#1}{(2\pi)^2}}
\newcommand{\ddt}[1]{\frac{d^3#1}{(2\pi)^3}}
\newcommand{\rr}{{\mathbf r}}
\newcommand{\kk}{{\mathbf k}}
\newcommand{\qq}{{\mathbf q}}

\begin{document}

\title{Intersubband polariton - LO phonon interaction in mid-infrared non-dispersive cavities: experimental demonstration of spontaneous scattering and perspectives towards polariton lasing}

\author{Paul Goulain\authormark{1} Mathieu Jeannin,\authormark{1}, Stefano Pirotta\authormark{1}, Adel Bousseksou\authormark{1}, Giorgio Biasiol\authormark{2}, Iacopo Carusotto\authormark{3}, Raffaele Colombelli\authormark{1} and Jean-Michel Manceau\authormark{1,*}}

\address{\authormark{1}Centre de Nanosciences et de Nanotechnologies, CNRS UMR 9001, Université Paris Saclay, 10 Boulevard Thomas Gobert, 91120 Palaiseau, France\\
\authormark{2}Laboratorio TASC, CNR-IOM, Area Science Park, S.S. 14 km 163.5, Basovizza I-34149 Trieste, Italy\\
\authormark{3}Pitaevskii BEC Center, CNR-INO and Dipartimento di Fisica, Università di Trento, I-38123 Trento, Italy}

\email{\authormark{*}jean-michel.manceau@cnrs.fr} 


\begin{abstract*} 
We report experimental evidence of the interaction between intersubband polaritons and longitudinal optical phonons in non-dispersive mid-infrared cavities, under resonant optical injection. The light emission originating from spontaneous polariton-phonon scattering  is observed at a frequency corresponding to an energy shift of one phonon below the pump frequency. Given the extremely low spontaneous scattering rate, we employ a custom-developed quantum mechanical model to numerically demonstrate the feasibility to stimulate such process using a pump–probe scheme. Based on this analysis, we identify a set of experimental conditions under which optical gain may be realized in a mid-infrared intersubband polaritonic system.
\end{abstract*}

\section{Introduction}

Over the past few decades, the precise control of light–matter interactions has evolved into a central theme in photonics and quantum electrodynamics. In the weak coupling regime, where the interaction strength between an emitter and a confined photonic mode remains below the dissipative rates, quantum electrodynamical effects such as the Purcell enhancement govern the modification of spontaneous emission rates. This regime enables efficient control over radiative processes and has been widely exploited in systems ranging from atomic cavities to solid-state quantum emitters \cite{goyObservation1983,gerardEnhanced1998}. As the coupling strength increases and surpasses the combined loss rates of the emitter and the cavity, the system enters the strong coupling regime, characterized by the reversible and coherent exchange of excitations between light and matter. This regime is marked by the emergence of hybridized excitation eigenmodes  —polaritons— exhibiting normal-mode splitting in the spectral domain, with the Rabi frequency setting the energy scale of the interaction. Strong coupling has been observed across a broad spectrum of platforms~\cite{raimond2001manipulating,kavokin2017microcavities,blais2021circuit}, with solid-state systems offering particular versatility due to their integrability and compatibility with advanced nanofabrication. Progress in lithographic patterning and epitaxial growth has enabled the fine-tuning of both photonic resonators and quantum emitters, facilitating systematic access to this non-perturbative regime \cite{basovPolariton2021a}.

In the mid- and far-infrared (MIR–FIR) spectral ranges, intersubband (ISB) transitions in semiconductor quantum wells constitute a particularly interesting platform for exploring strong coupling phenomena, owing to their large oscillator strengths and tunable transition energies. This system, which forms the backbone of quantum cascade lasers \cite{faistQuantum1994}, has proven especially useful for exploring quantum electrodynamical phenomena \cite{liuRabi1997,diniMicrocavity2003}. A key feature of ISB polaritons is that the  Rabi frequency is directly proportional to the population difference $\Delta n$ between the two subbands:

\begin{equation}
    \Omega_{\text{Rabi}}^2 = f_w \frac{f_{12}\Delta n ~e^2}{4 \varepsilon \varepsilon_0 m^{\ast} L_{\text{QW}}} \label{eq:Rabi}
\end{equation}
where $f_w$ is the ratio of the total quantum well length to the overall stack thickness, $f_{12}$ is the transition oscillator strength, $L_{\text{QW}}$ is the length of a single quantum well, and $m^*$ is the effective electron mass in the semiconductor under consideration. The population difference $\Delta n$ originates from the charge density introduced by the donor dopants: these are located in 
either the well or the barrier and establish a two-dimensional electron gas (2DEG) within the well. This was early on identified as a tuning parameter with significant implications both from a fundamental and applied perspective. In practice, applying a bias allows the QWs to be depleted, enabling a variety of experiments where the Rabi frequency is electrically modulated. This modulation permits the dynamic control of the system, toggling it in and out of the strong coupling regime. Such control proves particularly useful for implementing amplitude or phase modulation on a continuous-wave (CW) carrier at GHz frequencies, as demonstrated in \cite{anapparaElectrical2005,pirottaFast2021a,chungElectrical2023,malerbaUltrafast2024}. More recently, the same concept of entering and exiting the strong coupling regime has been proposed as a means to implement ultrafast saturable absorber mirrors, this time by optically saturating the transition \cite{julienOptical1988,zanottoUltrafast2012,mannUltrafast2021,jeanninUnified2021a,jeanninLow2023}.

ISB polaritons have also been the subject of more fundamental investigations, especially concerning their bosonic nature. A pioneering theoretical study of the quantum properties of such systems~\cite{ciutiQuantum2005} identified the possibility to increase the coupling strength way above the cavity and emitter decoherence rates up to the bare cavity and emitter frequencies and beyond~\cite{friskkockumUltrastrong2019a}. In this regime, the effects of the counter-rotating terms in the Hamiltonian—typically neglected—become significant and potentially observable. This so-called ultra-strong coupling regime holds the promise of extending quantum effects and technologies into the far-infrared (FIR) range of the spectrum, offering possibilities such as the emission of correlated photon pairs through abrupt modulation of the system, or the electrical probing of quantum phase transitions \cite{iqbalDynamical2024}. Demonstrating this regime has been the focus of significant efforts by the ISB community over the past decade \cite{todorovUltrastrong2010,geiserUltrastrong2012,dietzeUltrastrong2013}, with successful implementations across a variety of microcavity configurations and material systems \cite{askenaziUltrastrong2014,paulilloRoom2016b,jeanninUltrastrong2019,malerbaDetection2022b,goulainTHz2023,goulainScalable2024}. Equally intriguing are theoretical proposals leveraging the bosonic nature of these systems to reach final-state stimulation and inversionless lasing \cite{deliberatoStimulated2009}. Initial attempts to realize such devices through electrical carrier injection have thus far only yielded electroluminescence with relatively low efficiency \cite{geiserRoom2012,delteilOptical2011a,chastanetSurface2017a}. More recent approaches focus on resonant optical energy injection, exploiting interactions with longitudinal optical phonons or through parametric polariton processes \cite{manceauResonant2018c,knorrIntersubband2022a}.

The present work explores the possibility to obtain polariton-LO phonon scattering in non-dispersive microcavity. We experimentally measure the light emitted by the polaritons scattered via their interaction with LO-phonons. When tuning the pump frequency, we show that the emitted light is locked one LO-phonon below the pump frequency; we also show that we can take advantage of the non-dispersive nature of the photonic mode, changing the impinging angle of the pump light, leaving the emitted light frequency unchanged. We then numerically investigate the possibility to stimulate the scattering rate via the presence of a probe pulse, demonstrating the possibility to reach conditions where net gain could be observed.

\section{ Design, fabrication and characterization of the strongly coupled system}
\subsection{The multiple quantum wells stack}
To explore scattering processes in ISB polaritons, a quantum well stack was epitaxially grown. It consists of a multiple quantum well (QW) system: 7 repetitions of 8.3-nm-wide GaAs QWs separated by 20-nm-thick Al$_{0.33}$Ga$_{0.67}$As barriers. The nominal surface doping per QW ($n_{2D} = 4 \times 10^{12}~\text{cm}^{-2}$) is introduced as $\delta$-doping layers placed at the center of the barriers. The total thickness of the active region is 288 nm. A numerical simulation using a commercial Poisson–Schrödinger solver (NextNano software) is shown in Figure 1a.  The simulation includes the Hartree potential correction arising from ionized impurities in the barriers. The finite electron density in the two-dimensional electron gas then leads to a renormalization of the transition frequency, commonly referred to as the depolarization shift~\cite{zaluznyInfluence1993,todorovIntersubband2012,cominottiTheory2023}, given by the expression: $\widetilde\omega = \sqrt{\omega_{21}^{2} + \omega_{p}^{2}}$, where $\omega_{p}$ is the plasma frequency of the 2D electron gas. All together, the intersubband transition frequency $\widetilde\omega$ is expected to occur around 27~THz.

To evaluate the optical performance of the fabricated structures, the samples were shaped into a multi-pass prism, with facets polished at a 45$^{\circ}$ angle. A titanium/gold metallization was deposited on the top surface to enhance the overlap of the electric field with the QW stack. The samples were placed inside a Fourier Transform Infrared (FTIR) spectrometer that can be operated under vacuum. A polyethylene polarizer was used to select the polarization of the Globar source directed onto the sample. The transmitted spectra were recorded using a room-temperature Deuterated Triglycine Sulfate (DTGS) detector.

Due to the selection rules of intersubband transitions (ISBTs), the QWs selectively absorb transverse magnetic (TM) polarized light. To isolate this absorption, transmission spectra for TE- and TM-polarized incident light were measured, and their ratio was calculated. Figure~1b shows the resulting transmittance spectrum at room temperature, with a peak centered at 27.3~THz, in good agreement with the numerical simulation. From a Voigt profile fit, the linewidth of the transition is estimated to be 3~THz, corresponding to 10.9\% of the central frequency.


\begin{figure}[htbp]
\centering\includegraphics[width=10cm]{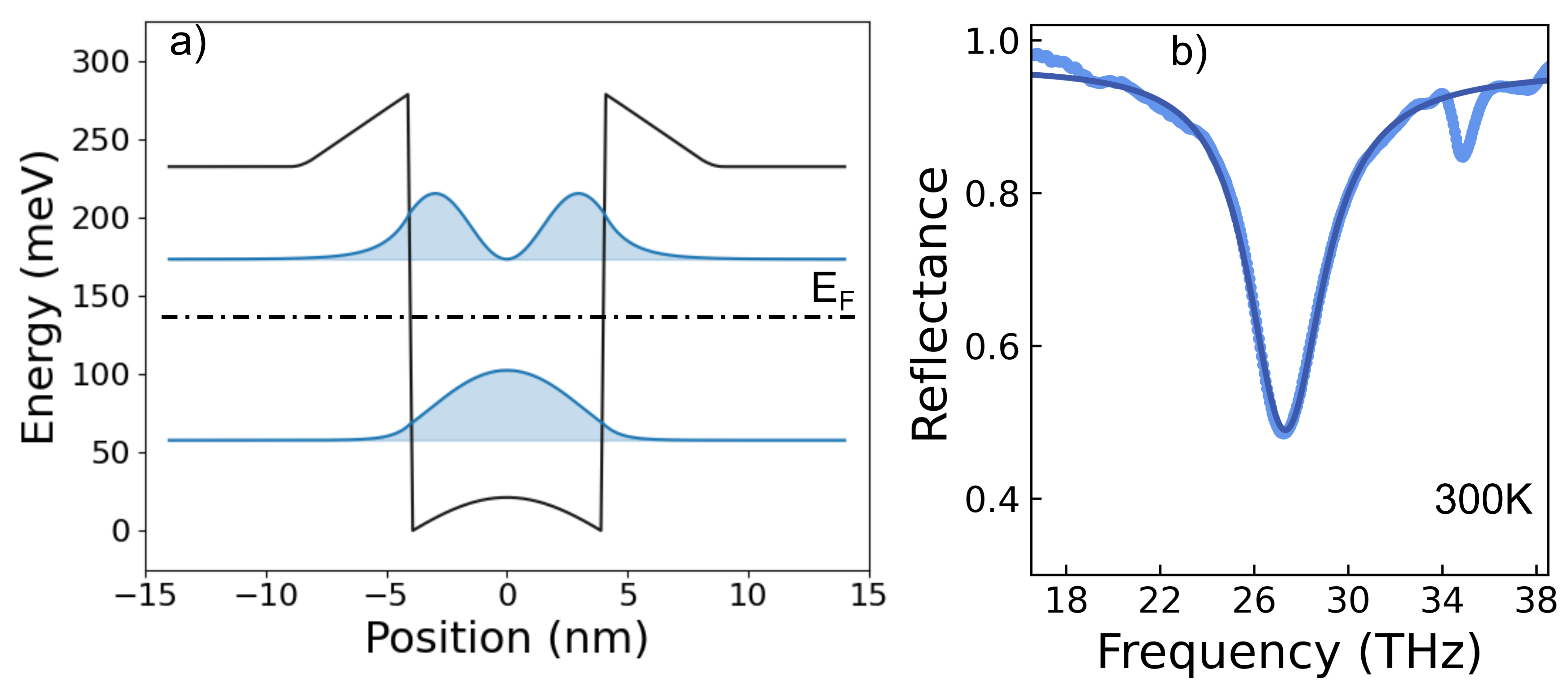}
\caption{(a) Simulated energy band profile and wavefunctions using a commercial Poisson-Schrodinger solver. (b) Transmittance of the sample shaped in a multipass prism configuration. The measurement is done at room temperature within an FTIR.}
\end{figure}

\subsection{Polaritons formation in the resonant optical micro-cavity}
Metal-Insulator-Metal (MIM) cavities are a class of photonic structures widely used in conjunction with mid-IR and THz quantum well (QW)-based devices due to their ability to confine electromagnetic modes with a non-zero electric field component in the z-direction—crucial for interacting with the confined two-dimensional electron gas in the QW stack. In these structures, waveguiding is achieved through boundary conditions imposed by the metallic layers, which restrict the propagation direction of light. MIM cavities have been extensively employed in intersubband physics for various purposes, including the formation of polaritons, the demonstration of the ultra-strong coupling regime, and integration with devices such as quantum well infrared photodetectors (QWIPs), modulators, and quantum cascade lasers (QCLs). The bottom mirror in these cavities is typically a flat gold layer, realized through thermo-compressive wafer bonding. Optical access to the microcavity is made possible by periodically patterning the top mirror along one direction (from now on called $x$) using lithographic techniques. This design offers a significant advantage by enabling surface probing of the system, thus facilitating a range of experimental techniques as discussed in  \cite{manceauMidinfrared2014a}. Depending on its thickness, the microcavity can operate in either a dispersive or non-dispersive regime. In the dispersive configuration, the photonic dispersion can be fully explored as a function of the incident angle. In contrast, in the non-dispersive regime, the cavity remains resonant at a fixed frequency regardless of the angle of incoming light along the $xz$ plane.

The latter is the case of interest for this work. In such a structure, the resonant mode is strictly confined below each metallic stripe, which will act as an independent cavity as extensively discussed in \cite{todorovOptical2010}. In the case of a 1D stripe with infinite length along $y$, the mode resonance frequencies depend on the stripe width and follow the formula:
\begin{equation}
    \nu_{M} = \frac{Mc}{2n_M\Lambda}
\end{equation}
where $M$ is the mode index, $n_M$ the effective refractive index of the mode, and $\Lambda$ the width of the stripe. Details of the fabrication process can be found in~\cite{manceauOptical2013a}.

In order to record the characteristic anticrossing behavior of polaritons, an array of microcavities with increasing stripe widths was fabricated, while maintaining a constant spacing of 1.4~$\mu$m between the stripes. A photograph of the sample is shown in inset of Figure~2a, where six different microcavity devices can be seen. Figure~2a presents the reflectance spectrum of the microcavity with a stripe width of 1.2~$\mu$m, measured using a Fourier Transform Infrared (FTIR) spectrometer equipped with a microscope. The reflectance was obtained by normalizing the sample's reflectivity to that of a flat gold mirror. A clear lifting of degeneracy is observed, with the appearance of new mixed eigenstates. By collecting together the measurements done on  each microcavity device, the polaritonic dispersion as a function of the cavity size can be reconstructed, revealing the classic anticrossing behavior of the hybrid states as shown on Figure 2b. From the minimum energy splitting, we infer the Rabi frequency of the system to be 2.65~THz.

\begin{figure}[htbp]
\centering\includegraphics[width=10cm]{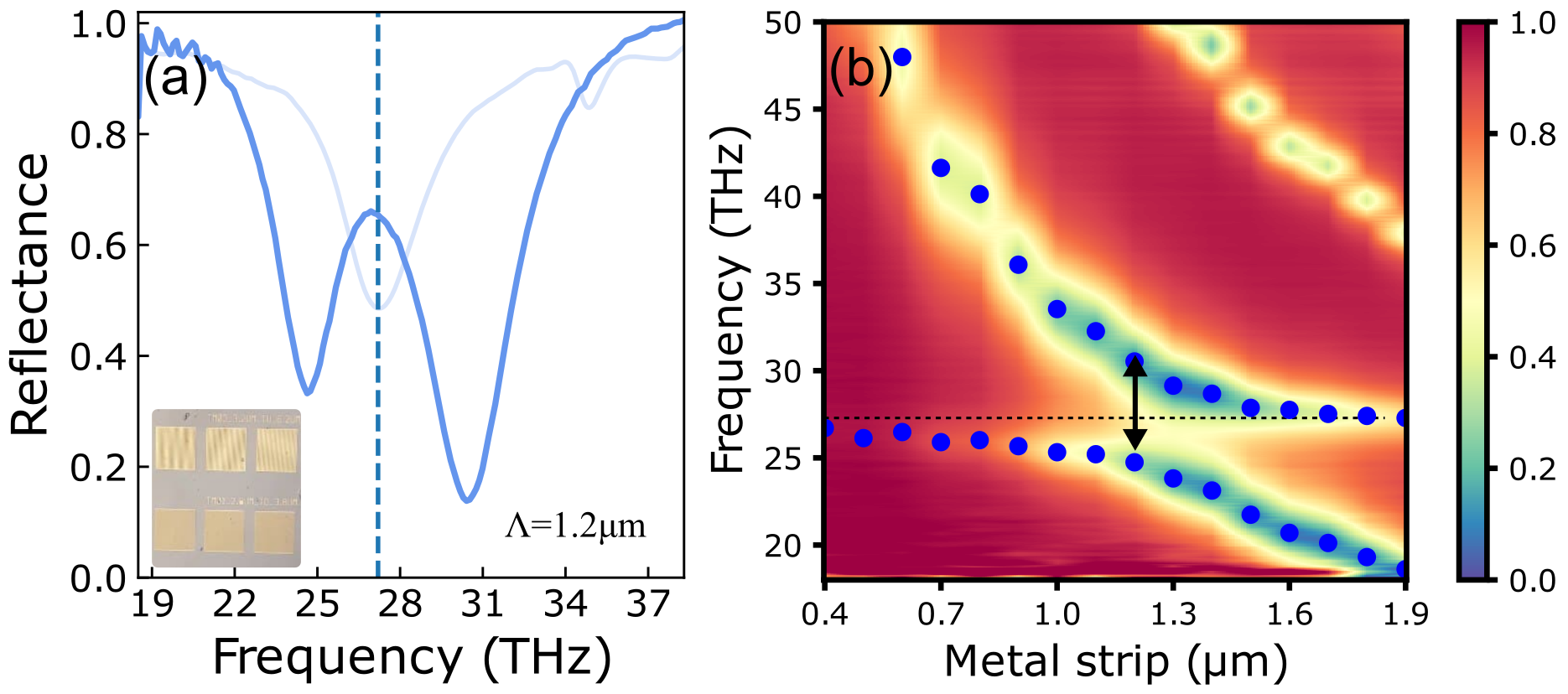}
\caption{(a) Reflectance measurement of the microcavity device with a stripe size of 1.2$\mu$m (the one closest to the anti-crossing point), along with the transmittance of the bare ISB transition (in transparency). The dash line marks the central frequency of the latter. In the inset, the optical image shows 6 microcavity arrays, each of them with a different metal stripe size. (b) Reconstructed experimental reflectance as function of the strip size. The arrow marks the position of the anti-crossing point within the polaritonic dispersion.}
\end{figure}

\section{ISB polariton - LO phonon scattering under resonant optical pumping condition}

In order to study the spontaneous phonon scattering processes under resonant light injection, we built an optical setup, presented in Figure~3a. The laser source is a commercial quantum cascade laser (QCL) from Daylight Solutions, tunable from 31 to 34.5~THz, with a maximum average output power in pulsed mode (100~kHz, 2~$\mu$s) of 110mW (the peak power within each pulse is five times higher). A compact goniometric arm directs the QCL output to precise positions in the energy-momentum space of the polaritonic system. The specular component of the reflected signal is measured using a power meter, enabling accurate alignment with the resonant polaritonic mode.

Using a beam profiler, we measured the beam waist at the focal point to be 170~$\mu$m (1/e$^{2}$), when focused with an anti-reflection (AR) coated ZnSe lens (f/\#2). The resulting average pump intensity can reach up to 750~W/cm$^{2}$ (3.75~kW/cm$^{2}$ peak power). The light emitted by the scattered polaritons is collected at normal incidence using an AR-coated ZnSe lens (f/\#1.5) and analyzed with a FTIR spectrometer coupled to a mercury cadmium telluride (MCT) detector cooled to 78~K. Long-wave pass filters are employed to reduce or extinct the amount of stray light scattered from the sample surface that reaches the detector.

\begin{figure}[htbp]
\centering\includegraphics[width=10cm]{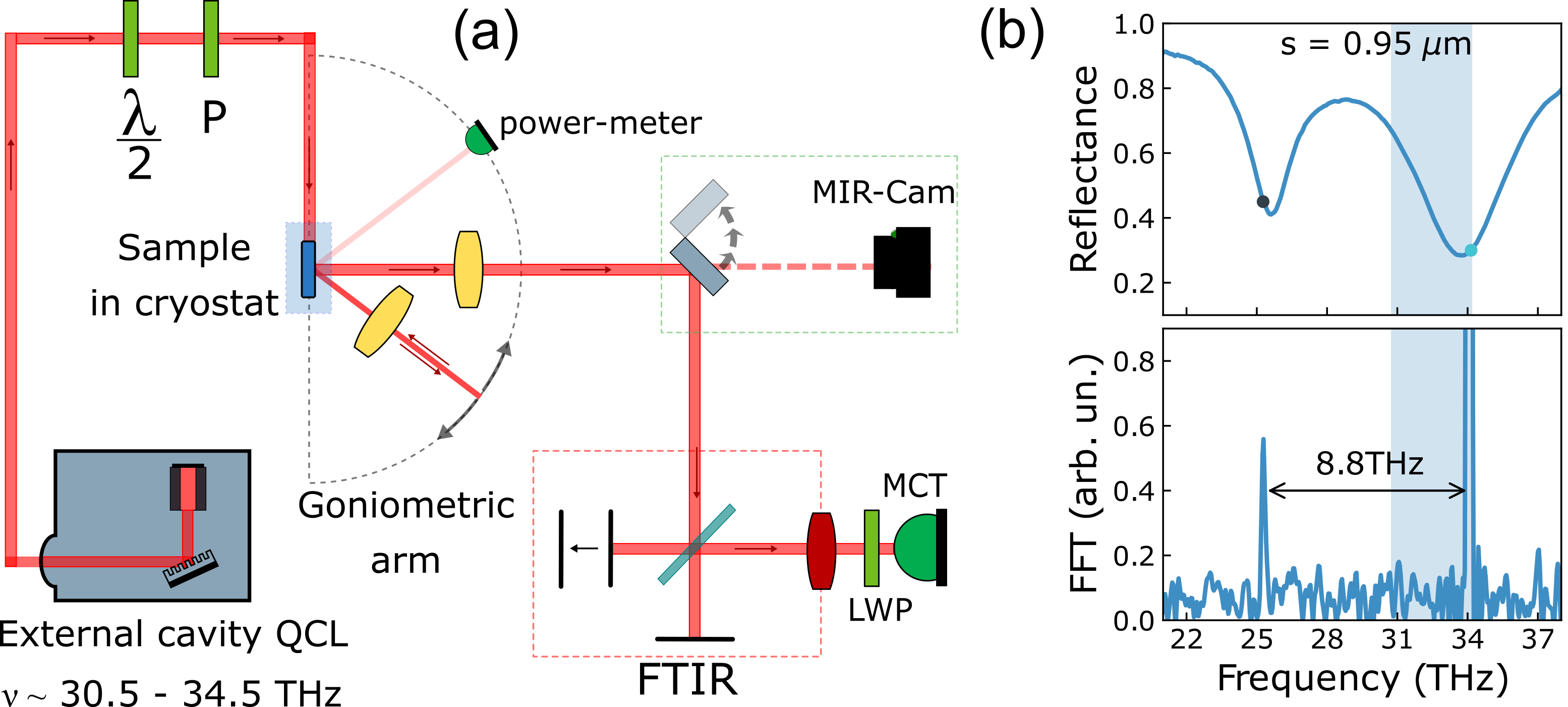}
\caption{(a) Experimental optical set-up with the tunable QCL and goniometric arms that allow the different experimental pump positions.(b) Upper panel, experimental reflectance of the microcavity with $\Lambda$ = 0.95 $\mu$m lateral size, recorded at 78K. The light blue dot corresponds to the pump state while the dark blue one is the expected final state. Lower panel, FFT of the recorded interferogram of the sample under resonant injection of the QCL at a frequency of  34.2 THz. The main peak is the remaining stray light from the pump. The lower peak corresponds to the photons emitted by the small fraction of polaritons scattered via LO-phonons. The blue shaded area is the QCL tuning range.}
\end{figure}

Our first experiment aimed to confirm the possibility of detecting light emission from polaritons that are scattered into a final state after emitting a longitudinal optical (LO) phonon. The upper panel of Figure~3b  displays the reflectance spectrum of the polaritonic system in a microcavity with a stripe width of 0.95~$\mu$m. This particular microcavity was selected because it provides an energy separation between the pumped and the final states of one LO phonon energy, while remaining within the tuning range of the QCL. The tuning range of the pump is indicated by the shaded blue area, and the positions of both the pump and the final states are marked with dots. Figure~3b (lower panel) shows the emission spectrum collected at normal incidence. The main peak at 34.2~THz corresponds to residual stray light from the pump (e.g. using only one filter). The lower-frequency peak originates from light emitted by the small fraction of polaritons that have scattered via LO phonons. The peak separation of 8.75~THz matches the LO phonon energy in GaAs.  

We further investigated this polaritonic configuration by tuning both the frequency and the in-plane momentum of the injected light. Figure 4a shows the numerically calculated polaritonic band-structure of the sample and the different positions in frequency and angle that have been investigated. The left panel of Figure 4b shows the Fourier transform of the recorded signal. The emitted light from the scattered polaritons remains strictly locked to the LO$_{ph}$ energy separation when varying the pump frequency from 34.2 THz down to 33.3 THz. 

More notably, when tuning the in-plane momentum of the injected polaritons along $x$, the signal remains locked to the LO$_{\text{ph}}$ energy, highlighting the non-dispersive nature of the cavity mode. One potential problem that arises from such non-dispersive devices is that the flat dispersion implies a continuum of possible final states at different $k_x$ values. This continuum of almost-degenerate final states implies that there is no preferential state for polaritons to scatter to. Hence, scattered polaritons are spread across a large part of the angular range, and thus the spontaneous emission is also spread across a wide range of angles. 

To test this hypothesis, a measurement was conducted using the same pumping angle, frequency, and power, but with a different collection lens: a 1-inch ZnSe lens instead of the previously used 2-inch one (while preserving the focal length), hence reducing the cone of collection of light by a factor 2. In such configuration, no signal could be detected. Furthermore, although our initial hypothesis was that flattening the dispersion of the final state might enhance the scattering efficiency, no significant improvement was observed. The conversion efficiency of the process remains on the order of $10^{-8}$, consistent with values reported for dispersive cavities~\cite{manceauResonant2018c}. This outcome underscores the importance of enabling the selection of a specific final state for polariton scattering, which could be achieved by adopting an experimental approach, where a probe pulse would populate the final state of choice. 

\begin{figure}[htbp]
\centering\includegraphics[width=11cm]{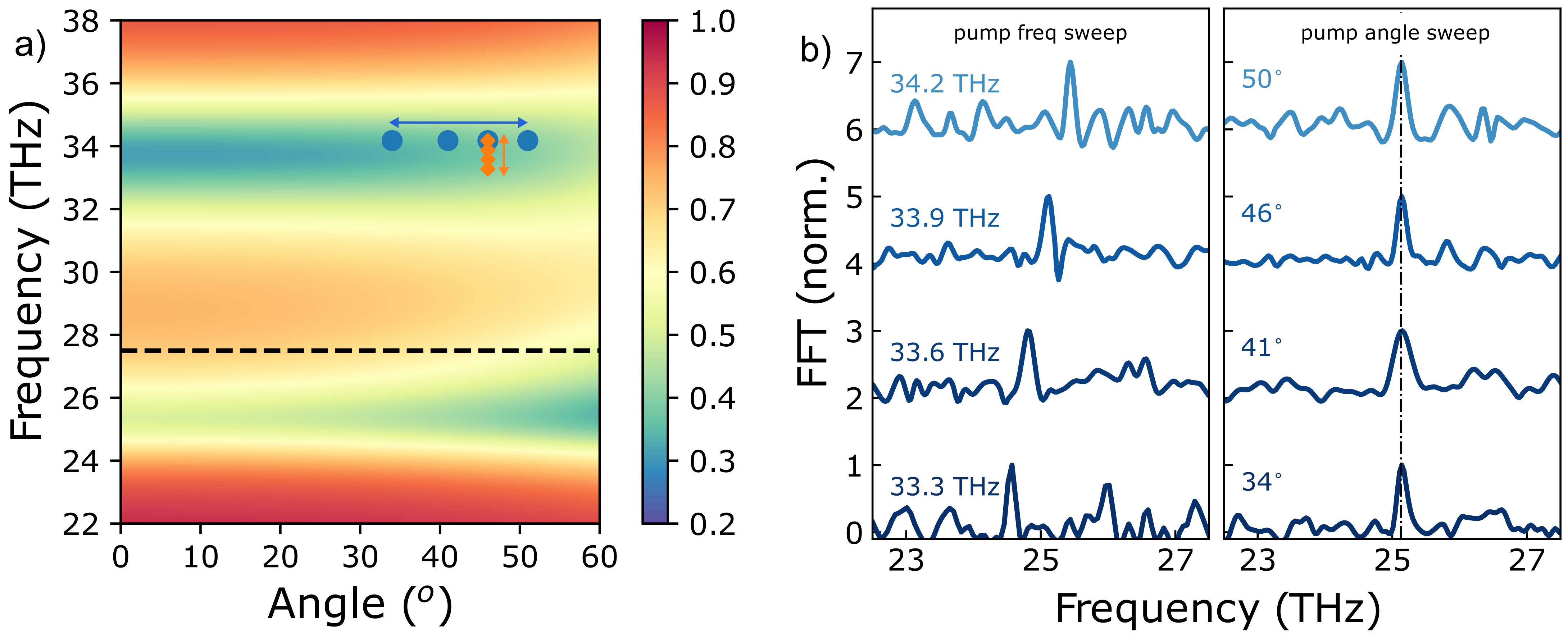}
\caption{(a) Numerically simulated transmittivity spectrum as a function of the incidence angle in the $xz$ plane. The different experimental pump positions are indicated. (b) Left panel, emission spectra when the pump frequency is changed while the angle is kept constant at 41$^{\circ}$. Right panel, emission spectra when the pump angle is changed while frequency is kept constant at 33.9 THz. The polaritonic spontaneous emission is locked 8.75 THz below the pump.}
\end{figure}

\section{Achieving gain conditions with a pulsed pump-probe configuration}
In stark contrast to fermions, bosons exhibit stimulated scattering, wherein the transition probability from an initial to a final state increases with the occupation of the final state. Specifically, if a scattering channel is available from a reservoir to a lower-energy state with a spontaneous rate $\Gamma_{0}$, the effective scattering rate can be significantly enhanced by the population density $n_{f}$ already present in the final state. This bosonic stimulation mechanism has been extensively harnessed in the exciton-polariton community, leading to the realization of non-equilibrium Bose-Einstein condensation \cite{baumberg2000parametric,PhysRevLett.85.3680,kasprzakBose2006,fontaineKardar2022,bloch2022non,st-jeanLasing2017}, opening the way to a variety of quantum many-body physics effects~\cite{carusottoQuantum2013} and offering an interesting platform for photonic quantum simulation \cite{amoExcitonpolaritons2016}.

As thoroughly described in \cite{ciutiQuantum2005}, ISB polaritons can be approximated as bosons in the dilute regime, making them subject to bosonic final-state stimulation and potentially enabling inversionless lasing \cite{deliberatoStimulated2009,colombelliPerspectives2015a}. While several demonstrations of electroluminescent devices operating at MIR and THz wavelengths have been reported, no threshold behavior indicative of stimulated coherent emission has been observed \cite{delteilOptical2011a,geiserRoom2012}. This is primarily because all devices developed to date have relied on resonant electrical injection into optically bright polaritonic states—a process that is extremely challenging. Even the use of narrow electrical injectors provides little improvement, as the majority of the electrical energy is still injected into the numerous electronic dark states of the system \cite{lagreeEffectivedensitymatrix2024}. This often results in thermally assisted electroluminescent devices \cite{chastanetSurface2017a}. Other promising schemes involve direct polariton-polariton scattering~\cite{nguyen-theEffective2013,nespolo2019generalized,cominottiTheory2023} or the polariton-phonon scattering process that are under examination in this work~\cite{deliberatoStimulated2009}.

In light of the parallels with exciton-polariton systems~\cite{savvidis2000angle}, a critical requirement for demonstrating stimulated scattering is the ability to populate the final state through a synchronized pump-probe experimental approach. In the case of ISB polaritons, such a regime was recently achieved using a non-collinear pump-probe geometry with phase-stable mid-IR pulses \cite{knorrIntersubband2022a}. The experimental observations were quantitatively recovered by a quantum-mechanical model of polariton-polariton scattering based on the Optical Bloch Equation formalism. In this work, we extend this quantum-mechanical model to include the LO phonon-polariton scattering channel and we numerically investigate the set of experimental conditions under which gain could be observed.

\subsection{Theoretical formalism}

In this Subsection, we start by introducing a general theoretical framework describing the coupling of intersubband polaritons to phonons. The readers that are not interested in the technical details can focus on the main results, namely the spatially local form of the phonon-polariton coupling Hamiltonian\eqref{eq:coupling_fin} and the expression \eqref{eq:coupling_constant_fin} for the coupling constant.

\subsubsection{Fr\"olich coupling between phonons and intersubband excitations}
Our theoretical formalism is based on the Fr\"olich coupling between electrons and phonons, which in a three-dimensional geometry has the standard form~\cite{moriElectronopticalphonon1989,ferreiraEvaluation1989}
\begin{multline}
 H_{e,ph}=\int\!\int\!\ddt{r}\,\ddt{q}\,\alpha(q)\,e^{-iqr}\, b_q^\dagger\, n_e(r) +\textrm{H.c.}= \\
 =\int\! \ddt{q}\, \alpha(q)\,\int\!\ddt{k}\,b^\dagger_q\, a^\dagger_k\, a_{k+q} +\textrm{H.c.}
 \label{eq:Froh}
\end{multline}
with a coupling constant
\begin{equation}
 \alpha(q)=\left[\frac{2\pi\hbar\omega_{LO} e^2}{\epsilon_\rho q^2}  \right]^{1/2}\,.
  \end{equation}
Here, $\omega_{LO}$ is the ($q$-independent) longitudinal-phonon frequency, the phononic contribution to the dielectric constant is written as $\epsilon^{-1}_\rho=[\epsilon_\infty^{-1}-\epsilon_s^{-1}]$ in terms of the static ($\epsilon_s$) and high-frequency ($\epsilon_\infty$) dielectric constants, and the electron density is
\begin{equation}
 n_e(r)=\int\!\ddt{k}\int\!\ddt{k'}\,e^{i(k'-k)r} \,a_k^\dagger\, a_{k'}
\end{equation}


In order to develop a two-dimensional formalism, we can express electron operators on the subband basis along $z$,
\begin{equation}
 a_{K,k_z}=\sum_j \tilde{\phi}_j(k_z)\,a_{K,j}
\end{equation}
where $j$ runs over the subbands of the quantum well potential of wavefunction $\phi_j(z)$ and 
\begin{equation}
\tilde{\phi}_j(k_z)=\int\!dz\,e^{-ik_z z}\,\phi_j(z)
\end{equation}
is the wavefunction of the $j$ subband expressed in reciprocal space.

Inserting this expansion into the Fr\"ohlich Hamiltonian \eqref{eq:Froh}, we obtain
\begin{equation}
 H^{2D}_{e,ph}
= \int\! \ddb{Q}\sum_{j,j'} C_{j,j'}(Q)\, \beta_{Q;j,j'}^\dagger\,\int\!  \ddb{K}\, a^\dagger_{K,j} a_{K+Q,j'}  +\textrm{H.c.} 
\label{eq:Froh_2D}
\end{equation}
where for each $j,j'$ pair and a given $Q$ the phonon creation operators are defined as suitable superpositions of $q_z$ modes matching the electronic transition,
\begin{equation}
\beta_{Q;j,j'}^\dagger = \frac{1}{ C_{j,j'}(Q)}\,\int\! \dd{q_z} \mathcal{I}_{j,j'}(q_z)\,\alpha(Q,q_z) 
 \,b^\dagger(Q,q_z)\,.
\end{equation}
The coupling coefficient $ C_{j,j'}(Q)$ is determined  by imposing Bose commutation rules 
\begin{equation}
[\beta_{Q;j,j'},\beta^\dagger_{Q';j,j'}]=(2\pi)^2\,\delta^{(2)}(Q-Q')
\end{equation}
for the phononic operators, which gives
\begin{multline}
 C_{j,j'}(Q)=\left[ \int\!\dd{q_z}\,\left| \mathcal{I}_{j,j'}(q_z)\,\alpha(q_z,Q)\right|^2 \right]^{1/2}= \\
 =\left[\frac{\pi\,\hbar\omega_{LO} e^2}{Q \epsilon_\rho} \,\int\!dz\,\int\!dz'\,e^{-Q|z-z'|}\,\phi_{j}^*(z)\,\phi_{j'}(z)\,\phi_{j'}^*(z')\,\phi_{j}(z') \right]^{1/2}
\end{multline}
in terms of the overlap factors 
\begin{equation}
 \mathcal{I}_{j,j'}(q_z)=\int \dd{k_z}\,\tilde{\phi}_j^*(k_z)\,\tilde{\phi}_j(k_z+q_z)=\int dz\,\phi^*_j(z)\,\phi_{j'}(z)\,e^{-iq_z z}\,.
\end{equation}

The Hamiltonian \eqref{eq:Froh_2D} has the simple physical interpretation of an electron scattering from the $j'$ to the $j$ subband while emitting a phonon of in-plane momentum $Q$ (and viceversa). In this work we are interested in phonon scattering processes involving intersubband polaritons that originate from the coupling of light to collective intersubband transitions. 

In terms of electronic operators, the creation operator of a ISB excitation from the first (1) to the second (2) subband has the form~\cite{todorovIntersubband2012}:
\begin{equation}
 B_k^\dagger = \frac{1}{\sqrt{n_{el}}}\,\int_{FS}\!\ddb{K}\,a_{K+k,2}^\dagger a_{K,1}
\end{equation}
where integration over $K$ is restricted to $|K|<k_F$ within the Fermi sphere and the sum over electron spin states is kept implicit for notational simplicity. We also restrict to a low-electron-density regime where the Coulomb interactions do not significantly distort the subband wavefunctions.
In the weak excitation limit where the density of ISB excitations on top of the filled Fermi sea in the first subband is much smaller than the electron density $n_{el}$, the ISB operators satisfy Bosonic commutation rules
\begin{equation}
[ B_k,B_{k'}^\dagger] = (2\pi)^2 \delta^{(2)}(k-k')\,.
\end{equation}
Physically, an ISB excitation at $k$ corresponds to a symmetric superposition of all possible electrons being promoted from $K$ to $K+k$. Thanks to this symmetry, the Fr\"ohlich Hamiltonian does not couple ISB excitations to the (non-symmetric) dark states.

In contrast to recent works focused on the (ultra)strong-coupling effects in the coherent interconversion between ISB polaritons and phonons and on the resulting Rabi splitting of ISB polarons modes~\cite{DeLiberato:PRB2012,ribeiroQuantum2020}, we are interested here in processes where ISB polaritons scatter via emitting/absorbing a phonon but the number of polaritons is not changed. Having such processes in mind and assuming the frequency of ISB excitations is well above the LO phonon frequency, we can restrict the sum in \eqref{eq:Froh_2D} to the $j=j'=1$ and $j=j'=2$ terms while neglecting the ones underlying the ISB plasmon-polaritons~\cite{DeLiberato:PRB2012,ribeiroQuantum2020} which are in our case off-resonance.

Within this approximation, the electronic term of \eqref{eq:Froh_2D} can be expressed in terms of ISB operators as
\begin{equation}
 H^{2D}_{e,ph}
= \int\! \ddb{Q} \left[C_{2,2}(Q)\, \beta_{Q;2,2}^\dagger-C_{1,1}(Q)\, \beta_{Q;1,1}^\dagger\right]\,\int\!  \ddb{K}\, B_K^\dagger B_{K+Q} +\textrm{H.c.} 
\label{eq:Froh_2D_ISB}
\end{equation}
where the two terms correspond to scattering of an electron in the second subband or of a hole in the first subband. This Hamiltonian can be further simplified into 
\begin{equation}
 H^{2D,pol}_{e,ph}
= \int\! \ddb{Q} \tilde{C}(Q)\,\tilde{\beta}_{Q}^\dagger\, \int\!  \ddb{K}\, B_K^\dagger B_{K+Q} +\textrm{H.c.} 
\label{eq:Froh_2D_ISB_fin}
\end{equation}
where the phonon creation operator $\tilde{\beta}_{Q}^\dagger$ includes the two processes with an overall coupling constant
\begin{equation}
|\tilde{C}(Q)|^2=\int\!\dd{q_f}\,|\alpha(Q,q_f)|^2\,
 |[\mathcal{I}_{2,2}(q_f)-\mathcal{I}_{1,1}(q_f)|^2
\end{equation}
that results from the sum of the amplitudes of the two processes. Inserting the explicit form of the Fr\"olich coupling into this expression gives
\begin{equation}
|\tilde{C}(Q)|^2=\frac{\pi\,\hbar\omega_{LO} e^2}{Q\,\epsilon_\rho}\,F(Q)
\end{equation}
where the details of the quantum well are summarized in the form factor \begin{equation}
    F(Q)=\int\! dz\int\! dz'\,e^{-Q\,|z-z|}\,\left[|\phi_1(z)|^2-|\phi_2(z)|^2\right]\left[|\phi_1(z')|^2-|\phi_2(z')|^2\right]\,.
    \end{equation}
For the typical processes under consideration, the phonon wavevector $Q$ is much smaller than the well thickness $L_w$, so the form factor can be expanded at lowest order in $Q$. This leads to an approximately $Q$-independent coupling constant
\begin{equation}
\kappa^2=\frac{\pi\,f\,\,\hbar\omega_{LO} e^2 L_w }{\epsilon_\rho}
\label{eq:coupling_constant_fin}
\end{equation}
where the coefficient, in the most relevant case of a square quantum well of thickness $L_w$ with infinite barriers, is $f\approx 0.032$.

The constant value of $\kappa$ allows for a straightforward reformulation of the Hamiltonian in real-space in the form
\begin{equation}
 H^{ISB-ph}=\kappa
 \,\int\!d^2\rr\,\Phi^\dagger(\rr)\,\Psi^\dagger(\rr)\,\Psi(\rr) + \textrm{H.c.}
 \label{eq:coupling_fin}
\end{equation}
where $\Phi(\rr)$ and $\Psi(\rr)$ are two-dimensional bosonic operators describing the phonon and the ISB excitation fields. These are coupled by a spatially local, nonlinear three-operator term describing parametric processes.
For realistic values of the system parameters such as $L_w=8.3\,\textrm{nm}$, $\epsilon_\infty=10.9$ and $\epsilon_s=12.9$ (giving $\epsilon_\rho\sim 70$), one gets $\kappa\sim 30\,\textrm{meV\,nm}$.

This interaction Hamiltonian is directly extended to an analogous coupling to ISB polaritons by including the Hopfield coefficient $u_{ISB}$ quantifying the ISB weight of the polariton. Note that, depending on the specific configuration at hand, some care might be required in including the momentum-dependence of $u_{ISB}$.

A straightforward Fermi golden rule calculation based on \eqref{eq:coupling_fin} provides a prediction
\begin{equation}
\Gamma_{sp}=
\frac{S_k}{2\pi \hbar}\,\kappa^2 \,|u^{in}_{ISB}|^2\,|u^{fin}_{ISB}|^2\,\frac{2}{\pi \hbar \Gamma_{pol}}
\end{equation}
for the spontaneous phonon emission rate by polaritons into an area $S_k$ of final states in momentum space, corresponding to a solid angle $d^2\Omega\approx S_k (c/\omega_{pol})^2$ for collection of scattered light. Given the relatively large value of the polariton linewidth $\Gamma_{pol}$, the weak dispersion of polariton states along the direction $y$ orthogonal to the patterning gives a negligible contribution. Plugging in specific numbers for the experimental configuration at hand and a collection cone with a radius of 5 degrees, one obtains a spontaneous scattering rate on the order of $5\cdot 10^4\,\textrm{s}^{-1}$, that is a factor around $10^{-8}$ smaller than the overall polariton decay rate. This value is comparable to the one predicted and measured in~\cite{manceauResonant2018c} for a dispersive cavity.

As a final comment, it is interesting to note how this phonon relaxation rate is relatively slow compared to the phonon-electron relaxation time, typically in the ps range. This is mostly due to the much lighter mass of the polariton modes (in dispersive configurations) or to the combination of a large polariton decay rate $\gamma_{pol}$ and the small momentum-space area of the relevant collected polaritons (in non-dispersive configurations).

\subsubsection{Generalized Optical Bloch Equations}

In order to obtain quantitative predictions in view of experiments for stimulated phonon-polariton scattering processes~\cite{deliberatoStimulated2009}, we can include the ISB-phonon coupling term \eqref{eq:coupling_fin} into the formalism developed in~\cite{knorrIntersubband2022a} for the description of the full nonlinear dynamics of ISB polaritons. Beyond the bosonic description, this also includes saturation and intensity-dependent shift of the ISB transition~\cite{andoElectronic1982a,zaluznyInterSubband1982a,zaluznyInfluence1993,julienOptical1988,todorovIntersubband2012,zanottoUltrafast2012,khurgin_coulomb_1991,craig1996undressing,batista2004rabi,Dietze_2013, Mann_Ultrafast_2021}.

This formalism is based on coupled differential equations for the cavity-photon field $E(\rr,t)$, the ISB coherence $\sigma(\rr,t)$, and the ISB population difference $\Pi(\rr,t)$, plus the additional classical field $\Phi(\rr,t)$ describing the phonons,
\begin{eqnarray}
    i\,\partial_t E &=& \left(\omega_{cav}-i\frac{\Gamma_{cav}}{2}\right) E  - \Omega_R \sigma + E_{ext}(\rr,t) \label{eq:OBE_E}\\
    i\,\partial_t \sigma &=& \omega_{ISB} \sigma + \Omega_R \Pi E + \eta(\Pi+1) \sigma -i \frac{\Gamma_{coh} 
    }{2}\sigma + \kappa (\Phi + \Phi^*) \sigma \label{eq:OBE_sigma}\\
    \partial_t \Pi &=& 2i\Omega_R (E \sigma^* - E^* \sigma) -\Gamma_{pop} (\Pi+1) \label{eq:OBE_Pi} \\
    i\,\partial_t \Phi &=& \omega_{phon} \Phi + \kappa n_{el}\,|\sigma|^2-i \frac{\Gamma_{phon}}{2} \Phi \,. \label{eq:OBE_phon}
\end{eqnarray}
The bosonic ISB excitation field $\Psi$ in \eqref{eq:coupling_fin} is related to the ISB coherence field $\sigma$ in the OBEs by $\Psi=\sqrt{n_{el}}\,\sigma$ where $n_{el}$ is the in-plane electronic density in the system.
As compared to the bosonic model of \eqref{eq:coupling_fin}, saturation of the ISB transition is included by the value of the population raising above the ground state value $\Pi=-1$: this provides a reduction of the Rabi frequency in \eqref{eq:OBE_sigma}. In this same equation, and a population-dependent depolarization shift is also visible, as discussed in~\cite{cominottiTheory2023} 

Here, $\omega_{cav,phon,ISB}$ are the bare frequencies of the cavity mode, of the LO phonon, and of the intersubband excitation (including the depolarization shift). Depending on the specific cavity configuration, the cavity mode may display a marked dependence on the in-plane wavevector, that translates into suitable differential operators in real-space. $\Omega_R$ is the Rabi frequency of the coupling between light and the ISB transition. $E_{ext}(\rr,t)$ is the (suitably normalized) amplitude of the coherent incident field driving the system. $\eta$ is the maximum depolarization shift at linear regime, $\Gamma_{phon,pop,coh}$ are respectively the decay rates of the phonon mode, of the population, and of the coherence. 

In what follows we will focus on the case of spatially wide pump and probe beams, so we can project the partial differential equations (\ref{eq:OBE_E}-\ref{eq:OBE_phon}) into the few relevant optical modes at the pump $\kk_p$ and probe $\kk_s$ wavevectors, and on the corresponding phonon mode $\Phi_q$ at $\qq=\kk_p-\kk_s$,
\begin{eqnarray}\label{eq:motion}
    i\,\partial_t E_{s,p} &=& \left(\omega_{cav}^{s,p}-i\frac{\Gamma_{cav}}{2}\right) E_{s,p}  - \Omega_R \sigma_{s,p} + E_{ext}^{{s,p}}(t) \\
    i\,\partial_t \sigma_{s} &=& \omega_{ISB} \sigma_{s} + \Omega_R (\Pi_0 E_s + \Pi^*_{q} E_p) + \eta [(\Pi_0+1) \sigma_s + \Pi^*_q\sigma_p] -i \frac{\Gamma_{coh}}{2}\sigma_s + \kappa \Phi_q^*\, \sigma_{p} \label{eq:OBEsigmas}\\
    i\,\partial_t \sigma_{p} &=& \omega_{ISB} \sigma_{p} + \Omega_R (\Pi_0 E_p + \Pi_{q} E_s) + \eta[(\Pi_0+1) \sigma_p +\Pi_q \sigma_s ]-i \frac{\Gamma_{coh}}{2}\sigma_p + \kappa \Phi_q\, \sigma_{s} \label{eq:OBEsigmap}\\
    \partial_t \Pi_0 &=& 2i\Omega_R (E^*_s \sigma_s + E_p^* \sigma_p - E_s \sigma^*_s - E_p \sigma^*_p)  -\Gamma_{pop} (\Pi_0+1) \\
    \partial_t \Pi_q &=& 2i\Omega_R (E_p \sigma_s^* + E_s^* \sigma_p ) -\Gamma_{pop} \Pi_q \\ \label{eq:OBEphonon}
    i\,\partial_t \Phi_q &=& \omega_{phon} \Phi_q + \kappa n_{el}\,\sigma_s^* \sigma_p -i \frac{\Gamma_{phon}}{2} \Phi_q \,
\end{eqnarray}
Here, $E_{p,s}$ and $\sigma_{p,s}$ are the spatial Fourier components of the electric field amplitude and of the electronic coherence at respectively the pump and probe wavevectors, while $\Pi_{0,q}$ are the spatial Fourier component of the population $\Pi(\rr)$ at respectively zero wavevector and at wavevector $\qq$ (reality of $\Pi(\rr)$ imposes that $\Pi_{-q}=\Pi_q^*$). As in this work the parametric process involves the (stimulated) scattering of a polariton at $\kk_p$ into a polariton at $\kk_s$ plus a phonon at $\qq$, there is no third polariton mode involved in the process as it was instead the case in~\cite{knorrIntersubband2022a}.

\subsection{Analytical estimates}

Before proceeding with the numerical calculations, it is interesting to get some analytical insight on the feasibility of observing marked stimulation effects in the phonon scattering process. Looking at in particular at \eqref{eq:OBEsigmas} and \eqref{eq:OBEphonon} and keeping in mind that the phonons are typically long-lived excitations, one notices that the threshold for parametric oscillation is roughly set by the condition 
\begin{equation}
\kappa^2  n_{el}\,|u^{fin}_{ISB}|^2 \,|\sigma_p|^2=\frac{\hbar^2\Gamma^{fin}_{pol}\Gamma_{phon}}{4}
\label{eq:thresh}
\end{equation}
where the decay rate $\Gamma^{fin}_{pol}= |u^{fin}_{ISB}|^2 \Gamma_{coh}+|u^{fin}_{cav}|^2 \Gamma_{cav} $ of the final ISB polariton state is a weighted average of the cavity and ISB transition linewidths $\Gamma_{cav,coh}$ and 
$u^{in,fin}_{ISB(cav)}$ are the Hopfield coefficients quantifying the weight of ISB excitation (photon mode) on the initial and final polariton states. As usual in parametric amplifiers, gain gets significant when approaching the threshold.
In our case, the condition $|\sigma_p|^2\leq 1$ holding for any saturable emitter provides a general bound 
\begin{equation}
\kappa^2 n_{el}\,|u^{fin}_{ISB}|^2 > \frac{\hbar^2\Gamma_{pol}\Gamma_{phon}}{4}
\label{eq:against_saturation}
\end{equation}
on the possibility of observing a significant gain before the ISB transition gets saturated. Most interestingly, this formula suggests that higher electronic densities are favorable in making the ISB transition robust against saturation. This theory can be generalized to the case of many quantum wells coupled to the cavity: the reduction by a factor $\sqrt{N_{QW}}$ of the phonon-polariton coupling $\kappa$ due the delocalization of both the phonon and the ISB polaritons among the different wells is compensated by the increased total number of electrons (at fixed electron density per well $n_{el}$) which conspire in maintaining the same relation \eqref{eq:against_saturation} for any number of wells. Of course, having a larger $N_{QW}$ is favourable in view of increasing the Rabi splitting between the lower and the upper ISB polariton branches.
Finally, it is worth highlighting how the small decay of phonons is an important advantage compared to the purely polariton scattering processes of~\cite{knorrIntersubband2022a}: on one hand, a small $\Gamma_{phon}$ facilitates reaching the condition \eqref{eq:against_saturation} for gain; on the the other hand, the reduced linewidth of the parametric gain peak gives an easier recognizable feature in the reflection/transmission spectra. This latter prediction will be confirmed by the numerical simulations we are going to present in the next Subsection.

Using this same arguments, we can obtain~\cite{knorrIntersubband2022a} an estimate for the incident pump power that is needed to reach the lasing threshold. Under a weak saturation condition $|\sigma_p|,\Pi+1\ll 1$, we can assume a linear response of the ISB transition to incident light. Around critical coupling to the incident radiation, most of the incident power enters the cavity. It is then possible to relate the in-cavity ISB polariton density to the incident power via 
\begin{equation}
I_{inc}=\eta\, \hbar \omega^{in}_{pol}\,\Gamma^{in}_{pol}\,\bar{n}_{in}
\end{equation}
where $\bar{n}_{in}$ is the polariton density in the initial mode and $\eta$ is a coefficient of order 1 accounting for the details of the radiative coupling process. 
Combining this formula with the general condition for parametric oscillation \eqref{eq:thresh} and keeping in mind that the elctronic coherence is related to the polariton density via $n_{el} |\sigma_p|^2 \sim \bar{n}_{in}\,|u_{ISB}^{in}|^2$, we get to the final estimate for the threshold intensity
\begin{equation}
 I_{th}=\frac{\hbar^2 \Gamma_{phon} \Gamma_{pol}^{fin} \Gamma^{in}_{pol} \omega_{pol}^{in}\,\epsilon_\rho}{4\pi\omega_{LO} e^2 f L_{QW}\,|u_{ISB}^{in} u_{ISB}^{fin}|^2}.
\end{equation}
Inserting realistic values of the parameters for our experimental device, we obtain a quite optimistic value in the $50\,\textrm{kW}/\textrm{cm}^2$ range.

\subsection{Results of numerical simulations}
Our numerical simulations are based on a set of experimentally extracted parameters. The complete polaritonic band is implemented in the model to reflect the non-dispersive nature of the system described above. This band dispersion is obtained from Rigorous Coupled Wave Analysis (RCWA) simulations, which are fitted to experimental data, following the approach used in several of our previous works. The intersubband (ISB) transition frequency ($\omega_{ISB}$) is set to 27.3 THz, with a linewidth ($\Gamma_{coh}$) of 1.6 THz, consistent with state-of-the-art experimental values. The Rabi frequency ($\Omega_{R}$ = 3 THz) and the cavity linewidth (Q$\approx$20) are chosen to match the observed experimental behavior. The system is modeled at the anti-crossing point, where the two polaritonic modes form perfectly mixed light–matter states. For the longitudinal optical (LO) phonon mode in GaAs, we use parameters from Ref.\cite{lockwoodOptical2005}: a central frequency ($\omega_{phon}$) of 8.75 THz, a damping rate ($\Gamma_{phon}$) of 69.8 GHz. For the effective coupling constant $\kappa\sqrt{n_{el}}$, we calculate from \eqref{eq:coupling_constant_fin} a value of 0.81 THz  using the electron density $n_{el} = 1.4 \times 10^{12}~\text{cm}^{-2}$ extracted from the experimentally observed Rabi splitting. The nonlinear coefficient in the optical Bloch equations ($\eta$ = 5 THz) is set in close agreement with the value reported in Ref.\cite{knorrIntersubband2022a}.

\begin{figure}[htbp]
\centering\includegraphics[width=10cm]{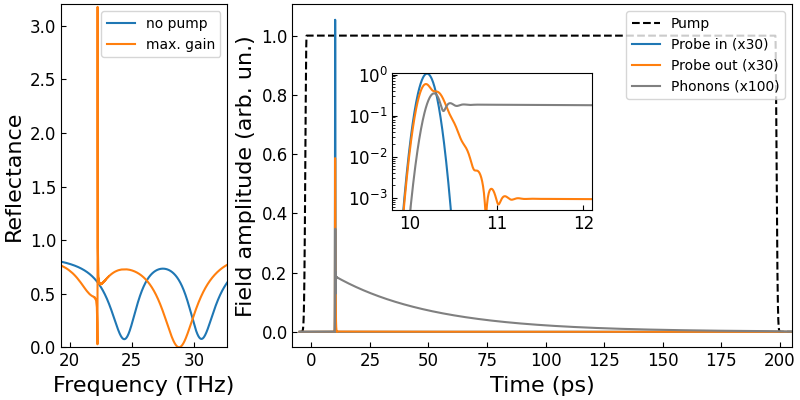}
\caption{(a) Numerically calculated reflectance of the sample at the conditions of no pump (blue) and of maximum observed gain (orange). (b) The absolute value of the pump field (dashed dark line), the probe field sent to probe the system (blue), the probe field after probing the system (orange), along with the phonon field (grey). The inset shows a temporal zoom of the fields in logarithmic scale.}
\end{figure}

To probe the final state and pump the ISB polariton reservoir, we define two optical pulses: a probe pulse modeled as an ultrashort Gaussian pulse with a duration of 70 fs, centered at 27 THz; and a pump pulse modeled as a quasi-continuous wave (quasi-CW) pulse centered on the upper polaritonic mode at 31 THz, with its duration optimized to enhance the stimulated scattering process. Due to the narrow linewidth of the LO phonon and the relatively low coupling efficiency, our numerical simulations indicate that the system reaches the stimulated regime only when using long pump pulses on the order of hundreds of picoseconds

Figure 5a shows the simulated reflectance of the polaritonic system both without the pump and with the optimal pump pulse parameters (power and duration, $\tau_{pump}=200$ ps) chosen to clearly reveal amplification. As expected, the amplified signal appears exactly 8.75 THz below the pump frequency, and the absolute reflectance reaches a value three times greater than unity, indicating the presence of optical gain. To further analyze this effect, Figure 5b presents the temporal evolution of the probe field before and after interacting with the sample under illumination, together with the generated coherent phonon field. The inset of Figure 5b shows in detail how the probe field is modified due to the stimulated scattering of the polariton population into the final state. Notably, there is a clear correspondence between the evolution of the probe field and the emergence of the coherent phonon field, as highlighted by the temporal zoom shown in the inset.

\begin{figure}[htbp]
\centering\includegraphics[width=10cm]{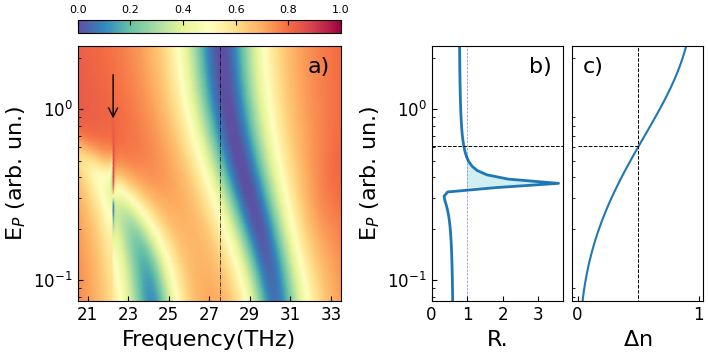}
\caption{(a) Simulated Reflectance of the probe pulse as a function of the energy of the pump pulse and of the probe frequency. The arrow indicates the frequency at which the gain (R>1) appears on the probe. The dash line marks the bare cavity resonance. (b) Reflectance of the probe pulse at a frequency of 22.25 THz as a function of the energy of the pump pulse. The blue colored area below the curve shows the region where one can observe gain. (c) Population difference between the two subbands of the QW as function of the energy of the pump pulse. Dash lines indicates the value at which the transition is saturatred, having half of the population in the 2nd subband.}
\end{figure}

As shown in previous experimental work \cite{knorrIntersubband2022a}, the amplification process is closely linked to the saturation of the polaritonic system. As the ISB reservoir is pumped, the population difference —and thus the Rabi frequency— decreases, leading to a renormalization of the system. Consequently, we perform a full simulation in which the pump pulse energy is gradually increased. Figure 5a presents a colormap of the sample reflectance as a function of pump pulse energy. At low pulse energy, both polaritonic states are clearly visible, indicating that the system remains almost unperturbed. As the pump pulse energy increases, we observe a decrease in the Rabi frequency, accompanied by a strong redshift of the upper branch and a less pronounced one of the lower one. More interestingly, a sharp feature appears one LO phonon below the pump frequency, at 22.25 THz. Figure 5b shows the sample reflectance at this specific frequency, where a clear gain is observed, with the reflectance exceeding unity. In parallel, Figure 5c shows the population difference between the two subbands as a function of pump pulse energy. The dashed line marks the pump energy at which saturation is reached (i.e., $\Delta n = 0.5$), while transparency is reached at $\Delta n = 1$. The rapid decline of the gain feature is due to a combination of the Rabi frequency reduction and the redshift of the injection state, which in turn reduces the efficiently of the pump polariton injection process into the system.

\section{Conclusion}
In summary, this work has experimentally demonstrated the interaction between intersubband polaritons and LO phonons in non-dispersive MIM cavities. The spontaneous emission of polaritons that scattered with LO-phonons from the upper branch is relatively weak, primarily due to the combination of a large polariton decay rate ($\gamma_{pol}$) and the limited momentum-space area available for the collected polaritons. 
We have then numerically shown that implementing a pump–probe scheme could enable the system to reach the regime of final-state stimulation, where optical gain can build up on the probe pulse. 

Although this result is exploratory, the predicted conditions for observing stimulation and gain appear experimentally achievable, paving the way for the realization of inversionless lasing and potentially polariton condensation at infrared and THz wavelengths.

\section*{Acknowledgements}
This work was supported by the European Union Future and Emerging Technologies (FET) Grant No. 737017 (MIR-BOSE). This work was partially supported by the French RENATECH network. IC acknowledges financial support from the Provincia Autonoma di Trento, from the Q@TN Initiative, and from the National Quantum Science and Technology Institute through the PNRR MUR Project under Grant PE0000023-NQSTI, co-funded by the European Union - NextGeneration EU.

\section*{Disclosures}
The authors declare no conflicts of interest


\bibliography{z_Article_phono_scatt}
\end{document}